# Impact Of Internet Governance

Saraswati Mishra[#1], Shikha Dhankar[*2], Kavita Choudhary[#3]

[#1, *2] *Students, Department of EECE, ITM University*
*Gurgaon, INDIA*

[#3] *Asst. Prof., Department of CSE, ITM University*
*Gurgaon, INDIA*

*Abstract*— This paper represents the overview of the influence of internet governance in all the transactions, trading, business, social services, educational activities, research etc. occurring through the internet. It is very essential to implement laws and regulations in the field wherever the transaction takes place, either in the form of money, material or services. To avoid any kind of fraud and cheating and to establish a peaceful environment in physical as well as virtual word, it is essential to have some organization which assures safety and security. Here I-governance is the governing body that tries to take care of all those requirements. This paper gives the quick information about the emergence of the I-governance, its impacts and two levels of working i.e. at national level and an international level. We can conclude that it is necessary to follow some rules so that other's privacy is not hampered similarly if rules are broken then the convict must be punished.

## I. INTRODUCTION

Modernization and technological improvements have influenced our lives deeply. Today every task is easy to perform irrespective of its difficulty level, thanks to Internet for its incredibility. Learning new techniques, education, ticket booking, shopping, getting knowledge about tourist place, heath consultancy, business processes, latest event's information, communication etc. everything is possible through internet. The internet is a network of networks. Internet connection allows the user to communicate with any other terminal globally, having internet connection. It establishes the web like structure over the world and provides a medium to share information. As the technology becomes more popular it is often gets misused by some people to harm others, hence we need proper governance. Governance is an act of setting standards, making rules and regulations and deciding punishments if the rules are broken. Since internet has become the part of day to day life, it should be governed to ensure safety and security. To solve these issues eGovernance and iGovernance are working world wide.

E-Governance (governance through electronic media) deals with the delivery of government services to citizens, interaction with business and industry, citizen empowerment through access to information, efficient government management through internet. These results in less corruption, increased transparency, convenience, revenue growth, and cost reductions.

Internet Governance (iGovernance) works for the development and application of principles, norms, rules, decision-making procedures, and programmes by Governments, the private sector and civil society, in their respective roles, that shape the evolution and use of the internet with minimization of cyber crime to ensure complete security of data as well as life safety [1].

## II. EMERGENCE OF INTERNET GOVERNANCE

A. *History of Internet and Governance*:

The concept of networking came in 1950s when a message was sent over the ARPANET at University of California, Los Angeles. After that many small networks were established. ARPANET also developed set of protocols for internetworking. In 1982, Internet Protocol suite (TCP/IP) was standardized and concept of internet was introduced. The internet was commercialized in 1995, since that time it has had revolutionary impact on culture, business, education and daily lives. Increase in number of users led to the research of advanced networks to support high data rare transmission. Today internet is spread over an every sector.

Internet is globally distributed network. It is not governed by any specific country, institute or business [2]. In 1995 when internet was spreading all over the world, its functionality and structure was being managed by the specialist bodies of respective fields, such as the Internet Corporation for Assigned Names and Numbers (ICANN) oversees assignment of domain name and IP address and is governed by international board of directors. The International Telecommunication Union (ITU) look after issues relate to technical standards and development. The World Wide Web Consortium (W3C) develops technical standards and guidelines for World Wide Web. The National bodies administer the national domains. These bodies were only concerned with managing and improving the internet's functionality, not with the governance. Internet governance first emerged on the international stage as a significant issue during the World Summit on the Information Society (WSIS) process in 2003 and 2005 [2]. This event gave support to the multi-stakeholder model of internet governance and led to the establishment of the Internet Governance Forum (IGF) at the final WSIS meeting in Tunis [2].

ICANN's status as a private corporation under contract to the U.S. government created disagreement amongst other governments, especially Brazil, China, South Africa and some Arab states. No common agreement existed on the definition of what comprised Internet governance. After much controversial dispute, during which the US delegation refused to consider surrendering the U.S. control of the Root Zone file, participants approved on a negotiation to permit for wider international debate on the policy principles. They agreed to establish an Internet Governance Forum, to be convened by the United Nations Secretary General before the end of the second quarter of 2006. The Greek government volunteered to host the first meeting.

B.  *Forum of internet governance:*

Establishment of Internet Governance Forum (IGF) was formally announced by the United Nations Secretary-General in July 2006 [1]. IGF brings together all stakeholders in the Internet governance including governments of different countries, the private sector or civil society, the technical and academic community, on an equal basis. The first meeting of the IGF was held in October 2006 .The theme was 'Internet Governance for Development'. The IGF is not a decision making body, it set up several "dynamic coalitions" to work on key issues, such as privacy, open standards, and an initiative on the rights and responsibilities of internet users [1].

C.  *Benefits of IGF:*

1) Free flow of ideas, knowledge and information over the internet.
2) Resolve issues related to security of user or network.
3) Improving accessibility all over the word to help developing countries.
4) Better understanding of cultural and linguistic diversity.
5) Establish standards of exchange of information to avoid piracy.
6) Developing methods to keep personal and confidential data secrecy [1].
7) Tracking unauthorized access.
8) Finding out methods and policies for better internet performance and governance.

III. INTERNET GOVERNANCE

Internet governance works on two levels national and international.

A.  *International iGovernance*

It is comprised of all the leading bodies working on an internet related to different sectors like trade and business, engineering and technology, social welfare society, research and development etc to solve issues and establish standards, rules and regulations which should be followed. If it is broken then the punishment and penalties are decided for the convicts.

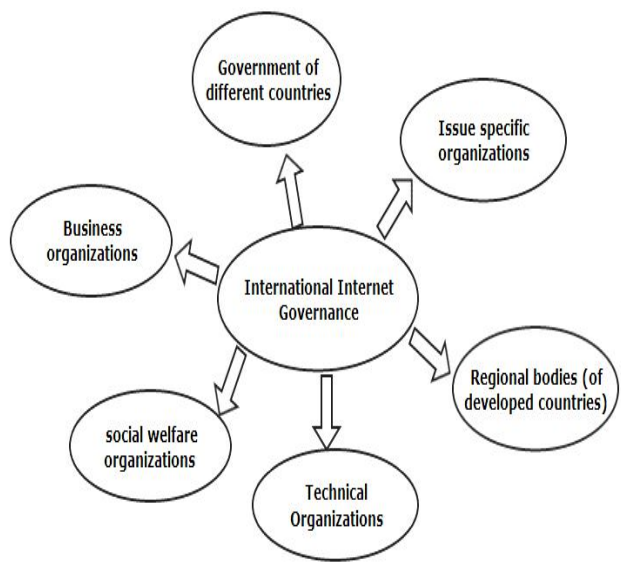

Fig.1 International level I-governance

B.  *Participation in iGovernance :*

There are some uncertainty regarding the participations of the countries and the distribution of power amongst them. Previously established group includes several bodies, with varying decision-making power. Regional bodies (U.S. government), issues Specific organizations (e.g. WTO), Human rights mechanisms (e.g. OHCHR), non-institutional international initiatives (e.g. Anti-Counterfeiting Trade Agreement (ACTA), the India, Brazil and South Africa dialogue (IBSA). This arrangement led to the conflicts in an organization due to unequal distribution of authority. The policy was found disadvantages for some stakeholders especially those who were from developing countries and want to contribute. Similarly much of the key policy decisions were taken by the private sector with no transparency or accountability.

Organizations are given power according to their contribution towards the network development and offered services. Technological empowerment has given the countries and bodies more authority and responsibility [2]. For example the internet economy is dominated by successful US companies like Apple, Google and Facebook, and by European and US based telecommunications companies [2]. This raises concerns among developing countries that the economic benefit of the internet is not shared equally. Developing countries are now working for a role in internet governance.

C.  *National iGovernance:*

It is similar to the International iGovernance but effective in national issues. It comprises of stakeholders, decision making body and the issues to be solved [3]. Here country has its own set of laws and procedures to solve the issues. Laws differ from nation to nation.

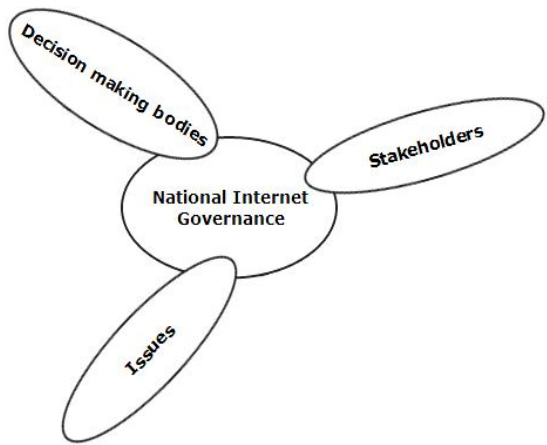

Fig. 2 iGovernance at national level

*1) Stakeholders in Internet Governance*

The stakeholders are the people who participate in discussion of the issues related to internet and communication and are affected by their outcomes [3]. Example - social bodies, business organizations, research department etc.

*2) Issues*

These are the problem related to communication, business activities, e-garbage, e-governance, health, education, data security, scam, cyber crime etc.

*3) Decision making Bodies*

The authorities of every field- information technology, communication technology, social, governmental, business, educational, research etc. are grouped together to find out finest and feasible solution of crisis [3].

IV. ISSUES ON WHICH IGOVERNANCE WORKS

Irrespective of national or international iGovernance, somehow they consistently try to solve same issues just their level and area of concern differ [4]. The issues are -
1) Agreements on standards and protocols for communication and networking [3] [4];
2) The structure of the domain name system and its management with the increase in number of systems;
3) Exchange of traffic between Internet Service Providers;
4) Act of branching on trademarks and intellectual property for companies or institutes having different classes and grades according to their services;
5) Economic, social and cultural issues–on development, rights and the environment [3][4];
6) Implementation of e-commerce and e-government;
7) Concerning taxation, cyber security or child protection;
8) Old problems such as spam and malware;
9) Recent developments such as social networking;
10) Innovations such as cloud computing have major impact on society.
11) Avoid misuse of new technologies like wireless sensor networks.

V. DECISION MAKING PROCESS

As the internet is a global medium, International Governance Forum provides a global discussion space in which issues of Internet governance is displayed and actual decisions are hidden. Final decision related to technical and communicational standard results from discussions and decisions in the International Telecommunication Union (ITU). Similarly, other expert organizations related to environmental, social, business, education, research and health etc. examine the open discussion forum and then implement their final decisions.

VI. RIGHT TO INFORMATION TO GET THE BENEFITS

Internet is available for all, every human being has right to get information about the internet policies, accessibility, regulations and government activities. Since the internet related concerns are increasing with each day, to ensure safety and security internet governance always guides to user for safe net surfing. Users can use their right to information to know more about cyber security, cyber law and government policies related to e-business. Various cyber offices are established to solve the cases of cyber crime, user should get their advantages.

VII. SOME LAWS ASSOCIATED WITH INTERNET GOVERNANCE

1) Technique for secure browsing that is filtering or blocking should be provided by the social networking sites.
2) Every user can create account on social networking site, use internet anywhere anytime.
3) Ban on the view and distribution of illegal Internet content such as child physical harassment and pornography.
4) Encouragement to specific religion hatred is a crime.
5) Creation of fake website is illegal.
6) Use of software which is not genuine is unethical.
7) Use of abusive words on social networking sites or anywhere else for a particular person, caste, religion, state or country is crime.
8) Material having copyright cannot be used without permission.

According to criminal law, penalties (fines or jail terms) will be charged on convict.

## VIII. FUTURE OF INTERNET GOVERNANCE

Some steps should be taken to flourish internet governance and build healthy environment all over the world.

1) Give equality to the governments of all the countries including developing countries in Internet Government Forum [5].
2) Empower the undersupplied countries to improve their communication facilities.
3) Establish a national database in each country for Internet resources to filter unwanted stuff.
4) Increase the awareness of the necessity of cleaning up rubbish information from the internet [5].
5) Provide proper information about the cyber laws and cyber crime to all internet users. (Specifically the age group 13 to 35).
6) Strengthen industry administration and regulate the information services market by law.
7) Support international exchange and cooperation and sharing of information and experience.

## IX. CONCLUSION

As there are two sides of coin, the technology facilitates the human beings at one hand and at another hand it brings various complex problems. It is our responsibility to use the advanced features for the benefit of the society not to harm others. When all the classes of the society will understand their rights and responsibilities, then there will be no crime, no harm, and no violence. This would be the crucial but necessary step to be taken by the iGovernance.